\documentclass[conference]{IEEEtran}
\IEEEoverridecommandlockouts
\usepackage{cite}
\usepackage{amsmath,amssymb,amsfonts}
\usepackage{algorithmic}
\usepackage{graphicx}
\usepackage{textcomp}
\usepackage{xcolor}

\usepackage{multirow}
\usepackage{hyperref}
\usepackage{siunitx}

\def\BibTeX{{\rm B\kern-.05em{\sc i\kern-.025em b}\kern-.08em
    T\kern-.1667em\lower.7ex\hbox{E}\kern-.125emX}}
\begin{document}

\title{Liver Segmentation in Time-resolved C-arm CT Volumes Reconstructed from Dynamic Perfusion Scans using Time Separation Technique \\
}

\author{\IEEEauthorblockN{Soumick~Chatterjee\textsuperscript{\textsection}\IEEEauthorrefmark{2}\IEEEauthorrefmark{3}\IEEEauthorrefmark{4}, 
                         Hana Haselji{\'c}\textsuperscript{\textsection}\IEEEauthorrefmark{1}\IEEEauthorrefmark{5}, 
                         Robert Frysch\IEEEauthorrefmark{1}\IEEEauthorrefmark{5},
                         Vojtěch Kulvait\IEEEauthorrefmark{6},\\
                         Vladimir Semshchikov\IEEEauthorrefmark{1},
                         Bennet Hensen\IEEEauthorrefmark{9}\IEEEauthorrefmark{5},
                         Frank Wacker\IEEEauthorrefmark{9}\IEEEauthorrefmark{5},
                         Inga Brüsch\IEEEauthorrefmark{10}\IEEEauthorrefmark{5},
                         Thomas Werncke\IEEEauthorrefmark{9}\IEEEauthorrefmark{5},\\
                         Oliver~Speck\IEEEauthorrefmark{4}\IEEEauthorrefmark{5}\IEEEauthorrefmark{7}\IEEEauthorrefmark{8},  
                         Andreas~N{\"u}rnberger\IEEEauthorrefmark{2}\IEEEauthorrefmark{3}\IEEEauthorrefmark{8} and 
                         Georg Rose\IEEEauthorrefmark{1}\IEEEauthorrefmark{5}\IEEEauthorrefmark{8}}
\\
\IEEEauthorblockA{\IEEEauthorrefmark{1}Institute for Medical Engineering, Otto von Guericke University Magdeburg, Germany}
\IEEEauthorblockA{\IEEEauthorrefmark{2}Faculty of Computer Science, Otto von Guericke University Magdeburg, Germany}
\IEEEauthorblockA{\IEEEauthorrefmark{3}Data and Knowledge Engineering Group, Otto von Guericke University Magdeburg, Germany}
\IEEEauthorblockA{\IEEEauthorrefmark{4}Biomedical Magnetic Resonance, Otto von Guericke University Magdeburg, Germany}
\IEEEauthorblockA{\IEEEauthorrefmark{5}Research Campus STIMULATE, Otto von Guericke University Magdeburg, Germany}
\IEEEauthorblockA{\IEEEauthorrefmark{6}Institute of Materials Physics, Helmholtz-Zentrum hereon, Geesthacht, Germany}
\IEEEauthorblockA{\IEEEauthorrefmark{7}German Center for Neurodegenerative Disease, Magdeburg, Germany}

\IEEEauthorblockA{\IEEEauthorrefmark{8}Center for Behavioral Brain Sciences, Magdeburg, Germany}

\IEEEauthorblockA{\IEEEauthorrefmark{9}Institute of Diagnostic and Interventional Radiology, Hannover Medical School, Hannover, Germany}

\IEEEauthorblockA{\IEEEauthorrefmark{10}Institute for Laboratory Animal Science, Hannover Medical School, Hannover, Germany}
}
\maketitle

\begingroup\renewcommand\thefootnote{\textsection}
\footnotetext{S. Chatterjee and H. Haselji{\'c} contributed equally}
\endgroup

\begin{abstract}
Perfusion imaging is a valuable tool for diagnosing and treatment planning for liver tumours. The time separation technique (TST) has been successfully used for modelling C-arm cone-beam computed tomography (CBCT) perfusion data. The reconstruction can be accompanied by the segmentation of the liver - for better visualisation and for generating comprehensive perfusion maps. Recently introduced Turbolift learning has been seen to perform well while working with TST reconstructions, but has not been explored for the time-resolved volumes (TRV) estimated out of TST reconstructions. The segmentation of the TRVs can be useful for tracking the movement of the liver over time. This research explores this possibility by training the multi-scale attention UNet of Turbolift learning at its third stage on the TRVs and shows the robustness of Turbolift learning since it can even work efficiently with the TRVs, resulting in a Dice score of 0.864±0.004. 

\end{abstract}


\section{Introduction}
Computed tomography (CT) perfusion imaging is a useful technique for diagnosing and planning therapy for liver tumours. With C-arm cone-beam CT (CBCT), the perfusion scan could be acquired in the interventional room and save time needed for patient transfer. For accurate visualisation and comprehensive liver perfusion maps, segmentations should be performed. However, liver segmentation in volumes reconstructed from the acquired CBCT perfusion scan is a difficult task due to several factors. The flow of the contrast agent affects the appearance of vessels. This, combined with noise, affects the grey values in the liver. The surrounding organs and tissue of the abdomen have similar attenuation values to those of the liver. 
The dynamic liver perfusion using C-arm CBCT is not yet explored enough, and not many acquisitions of different subjects are available. The C-arm rotates around the subject multiple times to acquire projections - which are then reconstructed to obtain the 3D volumetric images. This results in temporal undersampling and demands for a model-based reconstruction approach to be utilised.
So far, for liver perfusion, the time separation technique (TST)~\cite{Haseljic2021} emerges as the preferred choice over the straightforward reconstruction. Instead of treating all projections of one rotation to be acquired at the same time point (i.e. straightforward reconstruction), the TST models the temporal behaviour of every voxel as a linear combination of basis functions that are mutually orthonormal~\cite{Bannasch2018,Kulvait2022}. The liver segmentation was the subject of research in many works~\cite{Beichel2012-wq,Christ2016-zh,Draoua2014-ry}. However, none of these offered completely automated liver segmentation in CBCT volumes. Turbolift learning has been proposed~\cite{Chatterjee2022} to segment the liver from differently reconstructed CT and CBCT perfusion volumes, including TST.

Turbolift learning consists of three stages - CT, CBCT and CBCT TST. Each of the subsequent stages utilises the earlier stages as pretraining - helping to combat the problem of insufficient datasets is minimised. The first stage uses the reconstructed CT volumes for training, then the straightforward reconstructions of the CBCT volumes are used in the second stage, while the final stage uses the reconstructed coefficients by means of TST. However, neither the training nor the testing was performed using the reconstructed time-resolved volumes (TRV), estimated using the reconstructed coefficients. Segmentation of the TRVs can be utilised to visualise the change in the shape of the liver over time. Therefore, this paper analyses the segmentation results on the time-resolved volumes and if these results could be improved by working with time-resolved volumes in the third stage of the Turbolift instead of the reconstructed coefficients. This research replaces the third stage of the Turbolift with the TRVs and evaluates the robustness and applicability of Turbolift learning with TRVs. Segmenting the TRVs will deliver proper representation of the perfusion maps, and facilitate diagnosis and treatment of liver tumours.

\section{Methods}
Turbolift learning trains a multi-scale attention UNet in different stages, where the earlier training stages act as pretraining for the subsequent stage - assisting the model to learn from a small training dataset. The original Turbolift~\cite{Chatterjee2022} stages were CT, CBCT, and CBCT TST. Here, the CBCT TST implied the reconstructed first coefficient (FCR) by the means of TST. This was replaced here in this research by the time-resolved volumes (TRV), which were generated by all the reconstructed coefficients (shown in Fig.~\ref{fig:coeff}). This section starts by explaining the experimental setup and data acquisition methodology - explaining FCR and TRV in more detail, followed by a brief explanation of Turbolift learning, and finally exposing the implementation related information. 

\subsection{Experimental Setup and Data Acquisition}
The C-arm CBCT perfusion scans of four domestic pigs~\footnote{The research was carried out in compliance with European Directive 2010/63/EU and German animal welfare legislation (TierSchG). All experiments were approved by the local animal ethic committee (Lower Saxony State Office for Consumer Protection and Food Safety, LAVES 18/2809).} were acquired using Siemens Artis pheno C-arm using scanning protocol suggested in \cite{Datta2017}. It consists of ten rotations, with each covering rotation of \SI{200}{\degree} and with an angular step of \SI{0.8}{\degree}. The C-arm makes forward and backward rotations with a pause between every two successive rotations. The embolisation of the right hepatic liver artery was conducted in all four subjects resulting in decreased blood flow to embolised regions. 

\begin{figure}[htbp]
\centerline{\includegraphics[width=\columnwidth]{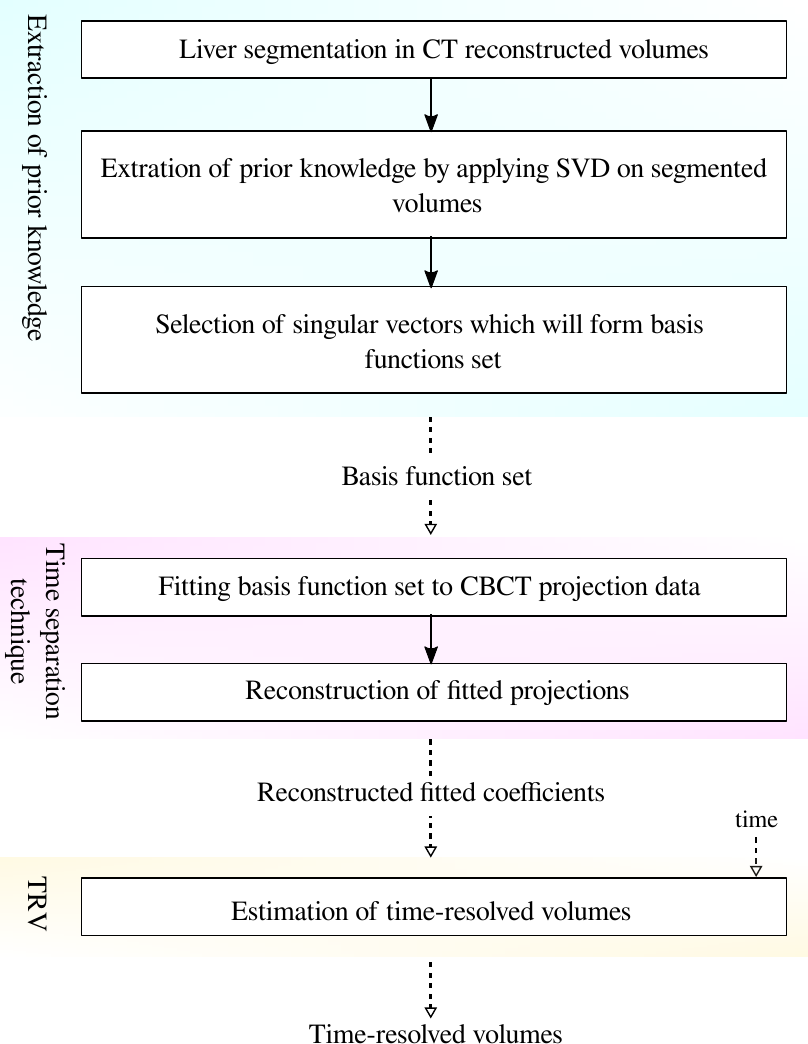}}
\caption{Steps for prior knowledge extraction and time separation technique application.}
\label{fig:flow}
\end{figure}

\begin{figure}[htbp]
\centerline{\includegraphics[width=\columnwidth]{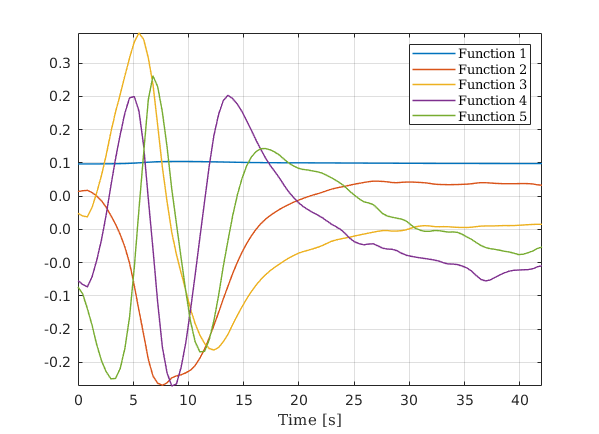}}
\caption{Basis function set used to model the C-arm Cone-Beam CT perfusion data.}
\label{fig:basis}
\end{figure}

The TST linearly combines orthonormal basis functions to model the change in grey values of every voxel and also of every pixel in projections during the scan duration. By multiplying both sides of the reconstruction problem $\mathbf{A x} = \mathbf{p}$ with any orthogonal basis function, the reconstruction problem is simplified. Instead of reconstructing a volume in every time point for which the projection was acquired during the scan duration, the number of reconstruction problems is the same as the number of basis functions. In this work, the basis function set has been constructed using prior knowledge, by applying singular value decomposition (SVD) on the reconstructed CT volumes, as shown in Fig.~\ref{fig:basis}. Note that the SVD was applied only to the liver regions, while the surrounding tissue, bones and catheters were ignored. The selected subsection of SVD vectors formed the basis function set, and the functions were fitted to the projections and were reconstructed (i.e. reconstructing fitted coefficients) with a slice thickness of 1mm. The basis set consisted of five basis functions. The analysis and discussion of the right number of SVD vectors is carried out in~\cite{Haseljic2022}. The general scheme of the described approach is depicted in ~Fig.\ref{fig:flow}.

\begin{figure}[htbp]
\centerline{\includegraphics[width=\columnwidth]{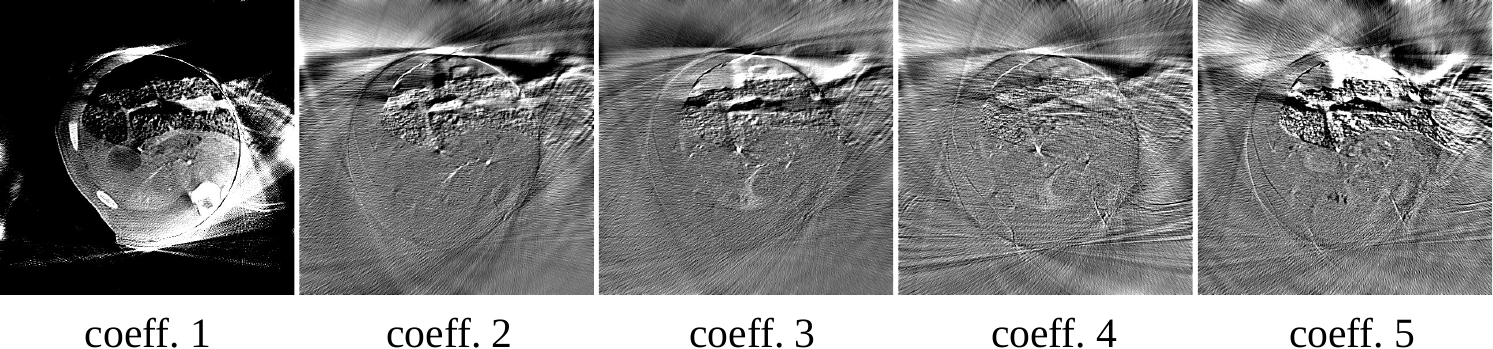}}
\caption{Reconstructed coefficients.}
\label{fig:coeff}
\end{figure}

Previous research~\cite{Chatterjee2022} only utilised the first coefficient for training since the associated basis function is a constant one and therefore the most similar to the classical reconstructed CT volumes (i.e. straightforward reconstructions). Fig.~\ref{fig:coeff} portrays all of the coefficients, and when analysed in relation to Fig.~\ref{fig:basis}, it is noticeable how the \textit{Function 2} contains information about the arterial input function and therefore models the contrast agent flow, but the liver contours are not easily distinguishable.

\begin{figure*}[htbp]
\centerline{\includegraphics[width=\textwidth]{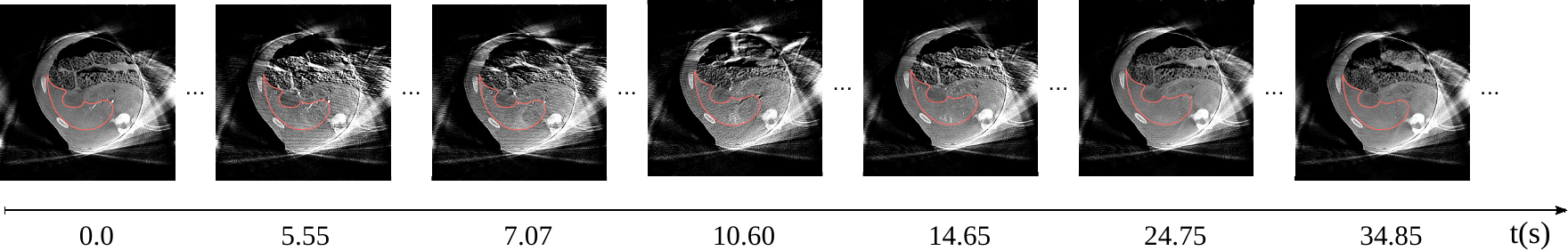}}
\caption{Time-resolved volumes reconstructed by the means of time separation technique. The liver is marked with red line. The change in contrast agent flow over time can be observed.}
\label{fig:trv}
\end{figure*}

In total, 100 samples of time-resolved volumes per subject were generated using these reconstructed coefficients over the complete scan interval. Since the contrast agent is washed out during scanning, many volumes appear very similar, as can be seen in Fig.~\ref{fig:trv}. Therefore, six volumes per subject were utilised in this research. In total, there were four subjects in the dataset - three of them were used for training and one for testing. In this manner, 4-fold cross-validation was performed (i.e. all possible combinations using those four subjects).

\subsection{Turbolift learning}
Deep learning typically requires large training datasets that are comparable to the final test set. Turbolift learning~\cite{Chatterjee2022} was proposed to combat this problem by employing a fine-tuning-based approach - using transfer learning in different stages. Turbolift starts by pretraining a modified version~\cite{kavur2021chaos,chaosMEMoRIAL} of the Multi-scale Attention UNet~\cite{abraham2019novel} on a dataset of healthy humans - the CHAOS dataset~\cite{kavur2021chaos}. The network has a UNet-like architecture~\cite{ronneberger2015u}, consisting of four encoding blocks, also known as the contraction path, to obtain the latent representation and then symmetrically uses four decoding blocks, known as the expansion path blocks, to obtain the final output. The output of each encoding block is supplied to its corresponding decoding block as skip connections. All skip connections, except the one in the final decoding step, pass through attention gates to suppress noise - making it an attention UNet~\cite{oktay2018attention}. Unlike typical deep learning models, this model receives the input image in the original scale and in three downsampled scales, which are supplied in the inner encoding blocks. Similarly, the output of the model is compared at different scales as well - known as deep supervision~\cite{zeng20173d} or multi-scale supervision~\cite{chatterjee2020ds6}. The final output of the model (output from the final decoding block), along with three more outputs from the inner decoding blocks, are compared against the original ground-truth (used in \cite{Chatterjee2022}), as well as three downscaled versions of the ground-truth, respectively. 

The CHAOS pretrained model was then trained using an animal CT perfusion dataset - constituting the first stage of the main training phase. The subsequent round of training of that trained model was carried out on the CBCT dataset before finally training on the first reconstructed coefficient of CBCT TST (FCR). In this stage-wise training method, the model learns on multiple datasets at different stages, allowing the model to perform well on the final and smallest dataset - FCR. The preceding training stages serve as pretraining for each subsequent task, allowing the model to benefit from three stages of pretraining (i.e. CHAOS pretraining and two stages of Turbolift) before finally learning on the FCR dataset. It has been seen that the performance of the model improves significantly from using Turbolift learning instead of training on the individual dataset~\cite{Chatterjee2022}. 

\begin{figure}[htbp]
\centerline{\includegraphics[width=\columnwidth]{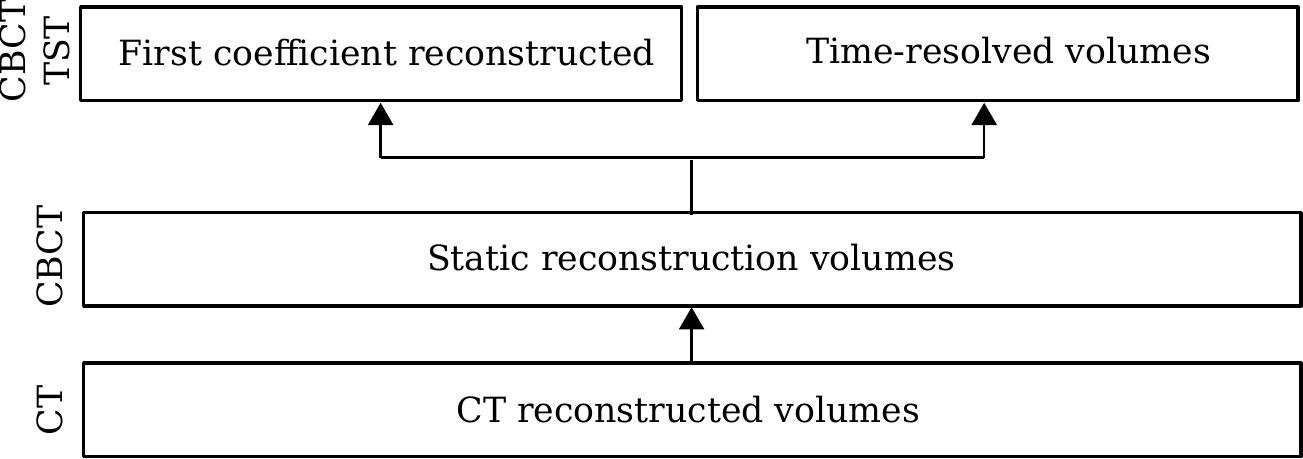}}
\caption{Three stages of Turbolift. In the last third stage, the training on reconstructed coefficient is replaced by training on the time-resolved volumes generated using this reconstructed coefficient.}
\label{fig:tlu}
\end{figure}

The earlier research of Turbolift learning~\cite{Chatterjee2022} works with the first reconstructed coefficient (FCR) of the CBCT TST reconstruction and not with the time-resolved volumes (TRV) estimated from it, as mentioned earlier, and segmenting the liver from these TRVs will be useful for tracking the movements of the liver over time. Hence, a second Turbolift was trained by replacing the final stage with TRV. In this case, the model trained on CBCT (already trained on CHAOS and CT) was trained on the TRV dataset. Fig.~\ref{fig:tlu} depicts the training process.  

\subsection{Implementation, Training, and Inference}
The original implementation of Turbolift~\footnote{Turbolift on GitHub:~\url{https://github.com/soumickmj/Turbolift}} was utilised in this research and was trained using 2D slices from the 3D volumetric images from the training datasets in various stages with a batch size of eight while collecting gradients of eight batches before backpropagating - resulting in an effective batch size of 64. The training was performed with the help of mixed precision~\cite{micikevicius2018mixed}. The error between the model's predictions and the ground-truth segmentation masks was measured using the focal Tversky loss~\cite{abraham2019novel} and was calculated at the original scale (the final output of the model) and three downsampled scales (as discussed earlier). All the four values were averaged to obtain the final loss value. This combined loss was then optimised for 500 epochs using the Adam optimiser with a learning rate of 0.001. The trained models were utilised for inference on that particular dataset at each stage of Turbolift, and then they were regarded as pre-training to conduct training on the next. 4-fold cross-validation taking three animals for training and one for validation, was performed at every stage of the training. To further aid the training on a small dataset, similar to the original work, four different data augmentation techniques were used with an overall probability of $75\%$ - random horizontal flips, random vertical flips, random rotation, and random translation of the pixels in both height and width directions. To remove noise from the predictions, the largest area in terms of the number of pixels was selected from each volume as there can only be one liver, and the rest was removed.

\section{Results and Discussion}

\begin{figure*}[pt]
\centerline{\includegraphics[width=\textwidth]{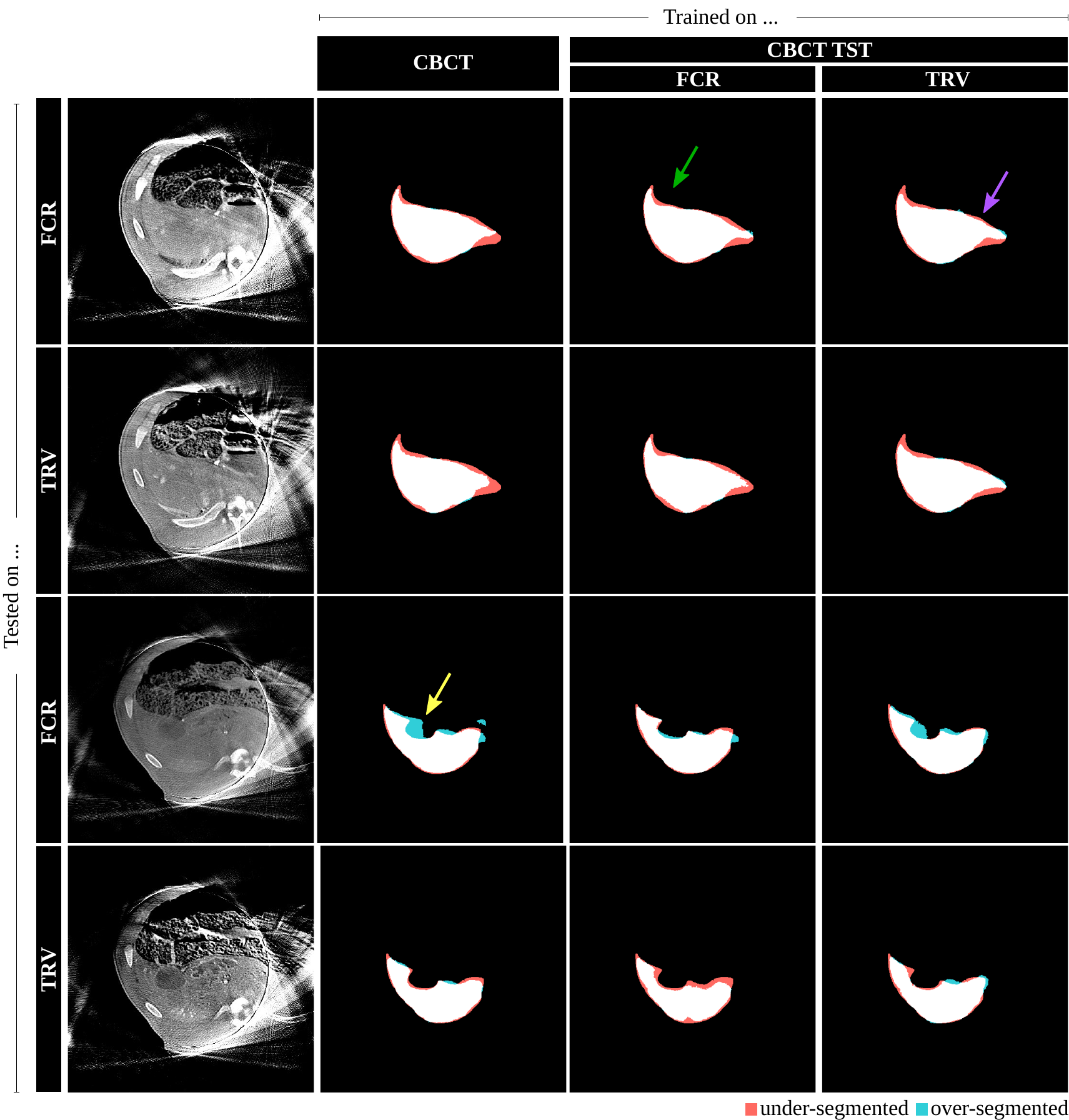}}
\caption{Qualitative comparisons of segmentations results of Turbolift CBCT stage (i.e. trained on straightforward reconstructions), Turbolift CBCT TST (i.e. training in third stage performed with reconstructed first coefficient) and Turbolift TRV TST (i.e. trained with time-resolved volumes) when tested on reconstructed first coefficients and time-resolved volumes.}
\label{fig:first}
\end{figure*}

Six sets of evaluations were performed during this research. For all of them, the first two stages of the Turbolift were as they were in the original paper~\cite{Chatterjee2022} - CT and CBCT. The third stage in the original Turbolift is referred to here as FCR (i.e. fitted coefficients reconstructed). One additional experiment was performed with the third stage using TRV (i.e. using the time-resolved volumes). For evaluation, the CBCT trained models (2nd stage of the Turbolift), FCR trained models (3rd stage of the original Turbolift), and TRV (3rd stage of the modified Turbolift) were evaluated using FCR and TRV. The results were evaluated using Dice and Intersection over Union (IoU), and are presented in Tables~\ref{results:dice}~and~\ref{results:iou}, respectively. 

It is worth mentioning that by modelling the data using only the selected subsets of the SVD components, not all the signals are modelled - which can also be considered as the denoising of the data. However, the presence of noise can be observed in functions in Fig.~\ref{fig:basis}. This means some noise components are also reconstructed and therefore might affect the segmentation quality. 

\begin{table}[htbp]
\caption{Dice of the segmentation results of traditional Turbolift learning (CBCT TST) and time-resolved Turbolift (TRV TST)}
\begin{center}
\begin{tabular}{c|ccc}
\hline
\multirow{2}{*}{\textbf{Trained on ...}} & \multicolumn{2}{c}{\textbf{Tested on ...}} \\
 & FCR & TRV \\ 
 \hline
 CBCT       & 0.890±0.019 & 0.865±0.087  \\ 
 FCR   & 0.905±0.007 & 0.842±0.141  \\ 
 TRV        & 0.883±0.003 & 0.864±0.004  \\ 
\hline

\end{tabular}
\label{results:dice}
\end{center}
\end{table}

\begin{table}[htbp]
\caption{Intersection Over Union (IoU) of the segmentation results of traditional Turbolift learning (CBCT TST) and time-resolved Turbolift (TRV TST)}
\begin{center}
\begin{tabular}{c|ccc}
\hline
\multirow{2}{*}{\textbf{Trained on ...}} & \multicolumn{2}{c}{\textbf{Tested on ...}} \\
 & FCR & TRV \\  
 \hline
 CBCT       & 0.802±0.019 & 0.762±0.076 \\ 
 FCR   & 0.826±0.010 & 0.726±0.124 \\ 
 TRV        & 0.790±0.006 & 0.761±0.007 \\ 
\hline

\end{tabular}
\label{results:iou}
\end{center}
\end{table}
\vspace{-10pt}





It can be observed that the method performs adequately in all scenarios as the lowest Dice score of 0.842±0.141 was obtained while testing TRV on the model trained with FCR, while the highest Dice score of 0.905±0.007 can be observed when FCR was tested on FCR-trained model. This trend is also confirmed by the IoU. Both metrics also show that when a similar type of data is used for training and testing, the results are better. Furthermore, it is noteworthy that the second stage of the Turbolift (CBCT) works better with TRV than the third stage (FCR) as the TRV is more similar to the volumes of straightforward reconstruction (CBCT) than to the reconstruction of the constant coefficient reconstruction. 

Fig.~\ref{fig:first} presents segmentation results for qualitative comparisons. The qualitative results corroborate the quantitative results - when the same type of data is used for training and testing, the results are better. The first set of examples (first two rows) shows the results are very similar in all six cases. However, the intestine region (green arrow) was better segmented when FCR trained model was tested on FCR data, while the stomach was better segmented by the model trained on TRV and tested on FCR (purple arrow). It is worth mentioning that the intensity values of the intestine are very different from the values of the liver. Even then, the TRV results in faulty segmentations, which might be due to the gallbladder region, which has similar values to the liver. But in general, both of them are better than CBCT results when tested on FCR and TRV. The second set of examples shows the region of the gallbladder. Similar to the earlier set, as well as the quantitative results, the models trained and tested on the same type of data perform better. When the FCR data was tested on CBCT and TRV trainings, the network failed to segment out the gallbladder properly (yellow arrow) as the gallbladder is not very well visible in the input FCR image. But, when it comes to TRV, the gallbladder is better visible and aids all three types of trainings in segmenting it. Considering the second stage of the Turbolift is trained on the straightforward reconstruction (CBCT) that experience change in the contrast agent flow over time, it will segment the TRV better than the FCR trainings. The FCR is only trained on the fitted reconstruction coefficient, which describes the static behaviour of an organ, but not the dynamic one (i.e. contrast flow).


\section{Conclusion}
This research presents an analysis of the segmentation quality of Turbolift learning while segmenting different types of computed tomography (CT) perfusion images. A multi-scale Attention UNet model was trained and tested serially on CT, straightforward cone-beam CT (CBCT) reconstruction, and first coefficient reconstruction (FCR) of model-based CBCT reconstruction TST - making the earlier stage act as the pretraining stage for the current one to combat with the problem of small datasets. In this research, the original Turbolift was modified by replacing the FCR stage with the time-resolved volume (TRV) data. The experiments revealed the robustness of Turbolift learning while encountering different scenarios, resulting in Dice scores between 0.842±0.141 and 0.905±0.007 while performing automatic segmentation. The experiments revealed that the method could work efficiently on both FCR, as well as TRV data. Further experiments will be performed to evaluate the performance of the method while encountering motion artefacts and while modelling using a different number of SVD components in TRV reconstruction.

\section*{Acknowledgement}
This work was in part conducted within the context of the International Graduate School MEMoRIAL at Otto von Guericke University (OVGU) Magdeburg, Germany, kindly supported by the European Structural and Investment Funds (ESF) under the programme "Sachsen-Anhalt WISSENSCHAFT Internationalisierung" (project no. ZS/2016/08/80646) and by the German Federal Ministry of Education and Research within the Research Campus STIMULATE (grant no. 13GW0473A and 13GW0473B).






\bibliographystyle{IEEEtran}  
\bibliography{ref} 

\begin{thebibliography}{10}
\providecommand{\url}[1]{#1}
\csname url@samestyle\endcsname
\providecommand{\newblock}{\relax}
\providecommand{\bibinfo}[2]{#2}
\providecommand{\BIBentrySTDinterwordspacing}{\spaceskip=0pt\relax}
\providecommand{\BIBentryALTinterwordstretchfactor}{4}
\providecommand{\BIBentryALTinterwordspacing}{\spaceskip=\fontdimen2\font plus
\BIBentryALTinterwordstretchfactor\fontdimen3\font minus
  \fontdimen4\font\relax}
\providecommand{\BIBforeignlanguage}[2]{{%
\expandafter\ifx\csname l@#1\endcsname\relax
\typeout{** WARNING: IEEEtran.bst: No hyphenation pattern has been}%
\typeout{** loaded for the language `#1'. Using the pattern for}%
\typeout{** the default language instead.}%
\else
\language=\csname l@#1\endcsname
\fi
#2}}
\providecommand{\BIBdecl}{\relax}
\BIBdecl

\bibitem{Haseljic2021}
H.~Haselji\'{c}, V.~Kulvait, R.~Frysch, B.~Hensen, F.~Wacker, G.~Rose, and
  T.~Werncke, ``{The Application of Time Separation Technique to Enhance C-arm
  CT Dynamic Liver Perfusion Imaging},'' in \emph{Proceedings of the 16th
  Virtual International Meeting on Fully 3D Image Reconstruction in Radiology
  and Nuclear Medicine}, 2021, pp. 264 --267.

\bibitem{Bannasch2018}
S.~Bannasch, R.~Frysch, T.~Pfeiffer, G.~Warnecke, and G.~Rose, ``Time
  separation technique: Accurate solution for {4D} {C}-arm-{CT} perfusion
  imaging using a temporal decomposition model,'' \emph{Medical Physics},
  vol.~45, no.~3, pp. 1080--1092, feb 2018.

\bibitem{Kulvait2022}
V.~Kulvait, P.~Hoelter, R.~Frysch, H.~Haseljić, A.~Doerfler, and G.~Rose, ``A
  novel use of time separation technique to improve flat detector ct perfusion
  imaging in stroke patients,'' \emph{Medical Physics}, vol.~49, no.~6, pp.
  3624--3637, 2022.

\bibitem{Beichel2012-wq}
R.~Beichel, A.~Bornik, C.~Bauer, and E.~Sorantin,
  ``\BIBforeignlanguage{en}{Liver segmentation in contrast enhanced {CT} data
  using graph cuts and interactive {3D} segmentation refinement methods},''
  \emph{\BIBforeignlanguage{en}{Med. Phys.}}, vol.~39, no.~3, pp. 1361--1373,
  Mar. 2012.

\bibitem{Christ2016-zh}
P.~F. Christ, M.~E.~A. Elshaer, F.~Ettlinger, S.~Tatavarty, M.~Bickel,
  P.~Bilic, M.~Rempfler, M.~Armbruster, F.~Hofmann, M.~D'Anastasi, W.~H.
  Sommer, S.-A. Ahmadi, and B.~H. Menze, ``Automatic liver and lesion
  segmentation in {CT} using cascaded fully convolutional neural networks and
  {3D} conditional random fields,'' in \emph{Medical Image Computing and
  {Computer-Assisted} Intervention -- {MICCAI} 2016}, ser. Lecture notes in
  computer science.\hskip 1em plus 0.5em minus 0.4em\relax Cham: Springer
  International Publishing, 2016, pp. 415--423.

\bibitem{Draoua2014-ry}
A.~Draoua, A.~Albouy-Kissi, A.~Vacavant, and V.~Sauvage, ``A new iterative
  method for liver segmentation from perfusion {CT} scans,'' in \emph{Medical
  Imaging 2014: Image Perception, Observer Performance, and Technology
  Assessment}, C.~R. Mello-Thoms and M.~A. Kupinski, Eds.\hskip 1em plus 0.5em
  minus 0.4em\relax SPIE, Mar. 2014.

\bibitem{Chatterjee2022}
H.~Haselji{\'c}, S.~Chatterjee, R.~Frysch, V.~Kulvait, V.~Semshchikov,
  B.~Hensen, F.~Wacker, I.~Br{\"u}sch, T.~Werncke, O.~Speck \emph{et~al.},
  ``{Liver Segmentation using Turbolift Learning for CT and Cone-beam C-arm
  Perfusion Imaging},'' \emph{arXiv preprint arXiv:2207.10167}, 2022.

\bibitem{Datta2017}
S.~Datta, K.~M\"{u}ller, T.~Moore, L.~Molvin, S.~Gehrisch, J.~Rosenberg,
  Y.~Saenz, M.~Manhart, Y.~Deuerling-Zheng, N.~Kothary, and R.~Fahrig,
  ``Dynamic measurement of arterial liver perfusion with an interventional
  c-arm system,'' \emph{Investigative Radiology}, vol.~52, no.~8, Aug. 2017.

\bibitem{Haseljic2022}
H.~Haselji\'{c}, V.~Kulvait, R.~Frysch, B.~Hensen, F.~Wacker, I.~Brüsch,
  T.~Werncke, G.~Rose, and F.~Sa'ad, ``Time separation technique using prior
  knowledge for dynamic liver perfusion imaging,'' in \emph{CT Meeting 2022},
  2022, \textit{accepted contribution}.

\bibitem{kavur2021chaos}
A.~E. Kavur, N.~S. Gezer, M.~Bar{\i}{\c{s}}, S.~Aslan, P.-H. Conze, V.~Groza,
  D.~D. Pham, S.~Chatterjee, P.~Ernst, S.~{\"O}zkan \emph{et~al.}, ``{CHAOS}
  challenge-combined {(CT-MR)} healthy abdominal organ segmentation,''
  \emph{Medical Image Analysis}, vol.~69, p. 101950, 2021.

\bibitem{chaosMEMoRIAL}
P.~Ernst, S.~Chatterjee, O.~Speck, and A.~Nürnberger, ``Chaos challenge - team
  ovgu memorial,'' 05 2019.

\bibitem{abraham2019novel}
N.~Abraham and N.~M. Khan, ``A novel focal tversky loss function with improved
  attention u-net for lesion segmentation,'' in \emph{2019 IEEE 16th
  international symposium on biomedical imaging (ISBI 2019)}.\hskip 1em plus
  0.5em minus 0.4em\relax IEEE, 2019, pp. 683--687.

\bibitem{ronneberger2015u}
O.~Ronneberger, P.~Fischer, and T.~Brox, ``U-net: Convolutional networks for
  biomedical image segmentation,'' in \emph{International Conference on Medical
  image computing and computer-assisted intervention}.\hskip 1em plus 0.5em
  minus 0.4em\relax Springer, 2015, pp. 234--241.

\bibitem{oktay2018attention}
O.~Oktay, J.~Schlemper, L.~L. Folgoc, M.~Lee, M.~Heinrich, K.~Misawa, K.~Mori,
  S.~McDonagh, N.~Y. Hammerla, B.~Kainz \emph{et~al.}, ``Attention u-net:
  Learning where to look for the pancreas,'' in \emph{MIDL 2018}, 2018.

\bibitem{zeng20173d}
G.~Zeng, X.~Yang, J.~Li, L.~Yu, P.-A. Heng, and G.~Zheng, ``3d u-net with
  multi-level deep supervision: fully automatic segmentation of proximal femur
  in 3d mr images,'' in \emph{International workshop on machine learning in
  medical imaging}.\hskip 1em plus 0.5em minus 0.4em\relax Springer, 2017, pp.
  274--282.

\bibitem{chatterjee2020ds6}
S.~Chatterjee, K.~Prabhu, M.~Pattadkal, G.~Bortsova, C.~Sarasaen, F.~Dubost,
  H.~Mattern, M.~de~Bruijne, O.~Speck, and A.~N{\"u}rnberger, ``{DS6,
  Deformation-Aware Semi-Supervised Learning: Application to Small Vessel
  Segmentation with Noisy Training Data},'' \emph{Journal of Imaging}, vol.~8,
  no.~10, p. 259, 2022.

\bibitem{micikevicius2018mixed}
P.~Micikevicius, S.~Narang, J.~Alben, G.~Diamos, E.~Elsen, D.~Garcia,
  B.~Ginsburg, M.~Houston, O.~Kuchaiev, G.~Venkatesh \emph{et~al.}, ``Mixed
  precision training,'' in \emph{International Conference on Learning
  Representations}, 2018.

\end{thebibliography}



\end{document}